\documentclass[twocolumn,showpacs,preprintnumbers,amsmath,amssymb]{revtex4}

\usepackage{epsfig}
\usepackage{graphicx}
\usepackage{dcolumn}
\usepackage{bm} 
\newcommand{\beq}{\begin{equation}}
\newcommand{\eeq}{\end{equation}}
\newcommand{\beqa}{\begin{eqnarray}}
\newcommand{\eeqa}{\end{eqnarray}}

\newcommand{\om}{\omega}

\def\oe#1{{ Opt.\ Express} {\bf#1}}

\def\jpb#1{{ J.\ Phys.\ B} {\bf#1}} 
\def\pra#1{{ Phys.\ Rev. A\/} {\bf#1}}

\def\prl#1{{ Phys.\ Rev.\ Lett.} {\bf#1}}

\begin{document}

\title{Double Ionization by Strong Elliptically Polarized Laser Pulses}

\author{Xu Wang}
\email{wangxu@pas.rochester.edu}
\author{J. H. Eberly}
\affiliation{ Rochester Theory Center and the Department of Physics 
\& Astronomy\\
University of Rochester, Rochester, New York 14627}


\date{\today}

\begin{abstract} We join the tribute to Professor N.B. Delone in this memorial issue by presenting the results of new calculations on the effects of ellipticity on double ionization by short and strong near-optical laser pulses. 
\end{abstract}

\pacs{32.80.Rm, 32.60.+i}

\maketitle


Observations of double ionization by femtosecond-scale near-optical laser pulses with intensities above about $10^{14}$W/cm$^2$ revealed a prominent surprising feature. The doubly ionized ion count, as a function of incident laser intensity, was observed to be significantly above the value predicted by the famous ADK ionization formula \cite{ADK}, which is used to describe non-correlated tunneling escape of two electrons in sequence. The presumption is then natural that a strongly correlated ionization channel has become open. The ADK formula and its predecessors \cite{Keldysh, PPT} are well matched by experimental facts for single ionization so it was startling to find them violated for high-field double ionization. Recent overviews of this domain of laser-atom physics are given by Agostini and DiMauro \cite{Agostini-DiMauro} and Becker and Rottke \cite{Becker-Rottke}.

A three-step  mechanism \cite{Corkum} has been generally accepted as responsible for highly correlated double ionization: first, one electron is freed with zero kinetic energy by tunneling through the coulomb barrier, which has been tipped by the laser field but is nearly static on the Bohr orbit time scale (Keldysh approximation \cite{Keldysh}); second, the freed electron is first pulled away and then driven back when the oscillating laser field reverses; and third, it acquires enough acceleration from the laser field to eject another electron in a near-core collision process. The same recollision process plays a key role in high harmonic generation \cite{Krause-etal}.

In this short note we present the first data from our theoretical analyses that include ellipticity of the laser field, suggesting this traditional view is too rigid, and may need modification. While the calculations themselves are not of the type that Professor Delone would have undertaken, as we will describe, we believe he would have been intrigued by the results emerging. The consequences of elliptical polarization, as an influence in high-field time-dependent atomic phenomena, are not at all trivial, but early reports mentioning tunnel ionization for elliptical polarization \cite{PPT, ADK} indicate that for cases not too close to circular one has a simple ratio of linear to elliptical tunnel rates that is independent of field intensity:
\beq
\frac{w_{ellip.}}{w_{lin.}} \approx \sqrt{\frac{1}{1-\varepsilon^2}},
\eeq
where $\varepsilon$ is the ellipticity. 

\begin{figure}[b!]
\includegraphics[height=6cm]{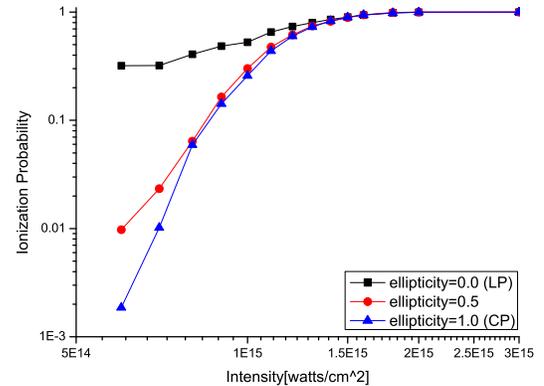}
\caption{{\footnotesize \label{fD.3curves} Predictions for double ionization as a function of laser intensity for ellipticity $\varepsilon$ = 0,\ 0.5,\ and 1.0.}}
\end{figure}

In Fig. \ref{fD.3curves} we show three double ionization probability curves as a function of $\varepsilon$. They indicate predictions of double ionization for $\varepsilon = 0$ (linear polarization), $\varepsilon = 0.5$, and $\varepsilon = 1$ (circular polarization). As one expects, the linear polarization curve remains a substantial fraction higher than the other two. However, it is intriguing that at the lower-intensity end of the curves all three deviate noticeably from the well-known smoothly and strongly decreasing curve typically provided by the ADK formula. We believe that this indicates an interesting domain for experimental work because such a deviation from the ADK curve is usually regarded as more or less impossible for finite ellipticity because the return collision in the three-step ionization mechanism is difficult to justify. 

Our approach, by which the curves in Fig. \ref{fD.3curves} were calculated, is heretical in the sense that it is fully classical, without even allowing tunneling to initiate the first electron's escape. However, a fully classical trajectory analysis \cite{classical} is well known for NSDI. In this alternative analysis two electrons can share energy inside the classical core potential and occasionally liberate one of them without tunneling. This picture has been satisfactory in explaining at all experimental double ionization (NSDI) features observed under linear polarization  \cite{classicalNSDI}, and has been extended to triple ionization \cite{triple} in good agreement with existing data, e.g., the relative shapes and strengths of momentum distributions in double ionization compared to triple ionization of argon \cite{triplecompare}.

Here we have used a two-dimensional ensemble \cite{classicalNSDI}, where the total energy of the system can be written as:
\beqa \label{e.EnergyEqnarray}
E_{tot} &=& \frac{1}{2}(P_{1x}^2 + P_{1y}^2 + P_{2x}^2 + P_{2y}^2)
\nonumber \\
&-& \frac{Z}{\sqrt{x_1^2 + y_1^2 + a^2}} - \frac{Z}{\sqrt{x_2^2 + y_2^2 
+ a^2}} \nonumber \\
&+& \frac{1}{\sqrt{(x_1-x_2)^2 + (y_1-y_2)^2 + b^2}}.
\eeqa
The first term is the total kinetic energy of the system, the second and third terms are soft-core Coulomb attractions \cite{classical} between the nucleus and the two electrons and the fourth term is the soft-core Coulomb repulsion between the two electrons. The size of the ensemble is chosen between 500K and 10M, depending on different ellipticities, which is large enough to get a statistically unbiased subensemble of double ionization events. We have set the total energy of each 2e member of the ensemble at -1.3 a.u., which is close enough to the ground state energies of both Kr and Xe to allow fairly close experimental relevance.

\begin{figure}[t!]
\includegraphics[width=8cm,height=5cm]{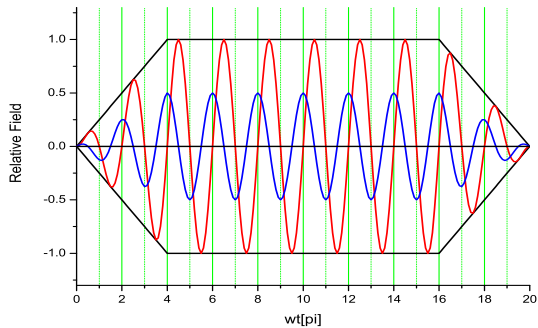}
\caption{{\footnotesize \label{fD.pulse} A trapezoidal pulse with phase $\phi=0$. The curve of higher amplitude is the $x$-component of the field and the lower curve is the $y$-component. The ellipticity is here taken to be $\varepsilon=0.5$. In calculations the phase $\phi$ in Eqn. (\ref{e.FieldEqn}) is set randomly to mimic the phase variation of repeated laser pulses.}}
\end{figure}

A many-pilot-atom method \cite{Abrines}, in the absence of a laser field, is used to generate a microcanonical zero-field ground state ensemble to start the interaction at t=0. A laser pulse as shown in Fig. \ref{fD.pulse} is used, where the linear two-cycle ramp has some theoretical advantages \cite{Grobe-Fedorov}. Approximately, it gives an iso-intensity interaction between field and electrons, but it sacrifices smoothing associated with more realistic pulses shapes.  

Once the microcanonical ensemble is generated, the laser field 
\beq \label{e.FieldEqn}
E(t) = E_0f(t)[\hat{e}_x\sin(\om t+\phi)+\hat{e}_y\varepsilon \cos(\om t + \phi)],
\eeq
is turned on. The circular frequency $\omega$ is set to be 0.0584 a.u., corresponding to 780 nm wavelength, consistent with most experiments using Ti:sapphire laser systems. A 10-cycle pulse with this frequency gives a pulse length of about 26 fs.  

\begin{figure*}[t!] 
\includegraphics[width=8cm]{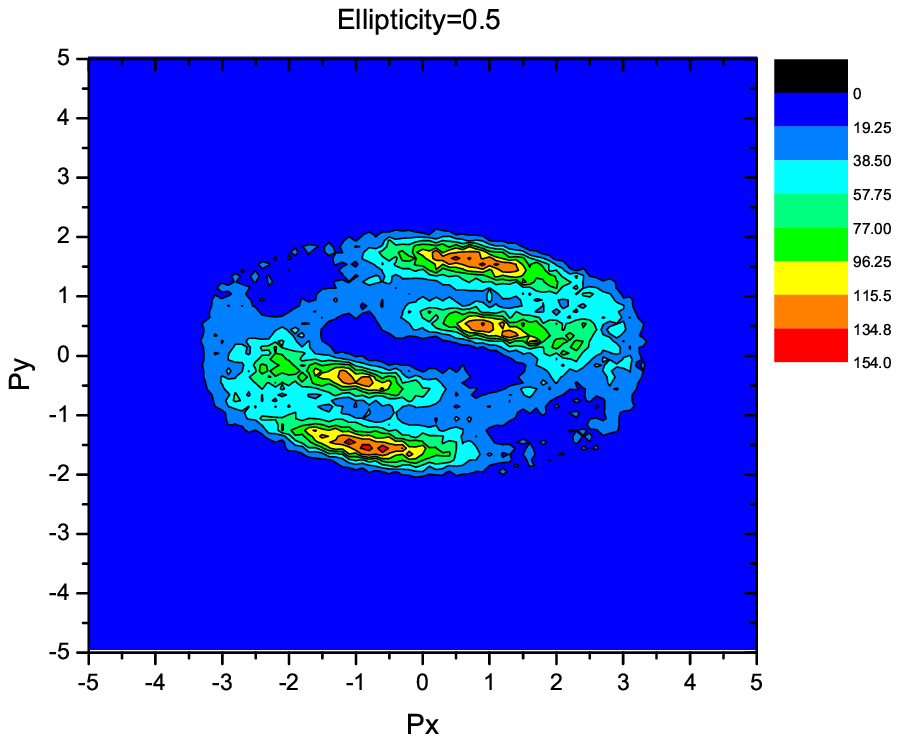} 
\includegraphics[width=7cm]{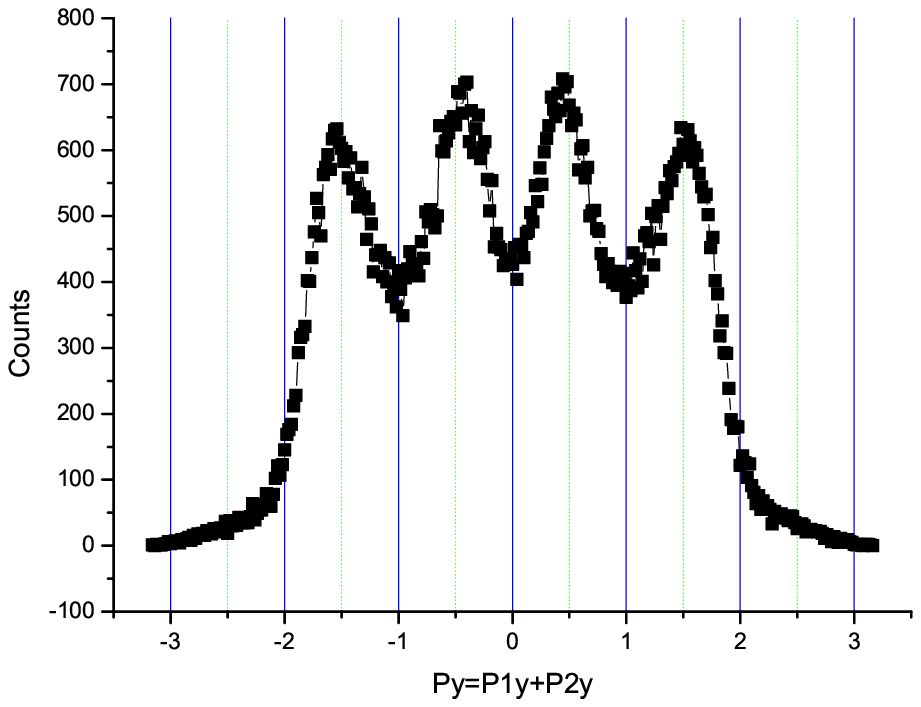}
\caption{{\footnotesize \label{fD.4bands}  Effects of elliptical polarization for $\varepsilon$ = 0.5. In the $p_x-p_y$ plot (Left) one sees a structured end-of-pulse momentum distribution for doubly ionized ions. A scan produces the $p_y$ distribution shown in the Right. }}
\end{figure*}

The classical potential's parameter values $a$  = 1.77 and $b$ = 0.1 serve to specify a model for the atomic core.  The amplitude of the laser field   is set to give a peak (not average) laser intensity of 0.6 PW. At the end of the laser pulse, the positions and momenta of the two electrons are recorded. The end-of-pulse ion counts are shown in Fig. \ref{fD.3curves} and the momenta of the doubly ionized ions, which are calculated as the sum of the momenta of the two ionized electrons, are shown in Fig. \ref{fD.4bands}(Left) as a distribution in the $p_x-p_y$ plane.

A scan of Fig. \ref{fD.4bands}(Left) over the distribution of ion momenta in the $x$-direction is a single-peak curve similar to what is seen with linear polarization. However, the $y$ direction is more interesting because it contains the consequences of non-zero $\varepsilon$. There are four rather sharp peaks of momenta, as shown in Fig. \ref{fD.4bands}(Right). These are features not available previously, and they appear to deserve careful future attention. 

Experience has shown that experiments in the PW regime of intensities and multi-femtosecond pulse durations provide detailed information about the double ionization process, particularly by examining the recoil ion momentum distribution. For elliptic or circular polarization, because of the dramatic drop of the probability of the first-freed electron to return directly to the ionic core, high-correlation ionization has not been detected in most rare gas atoms or molecules. Of course, one cannot automatically conclude that there is no NSDI for elliptic or circular polarization, and in fact characteristic knee signatures have been observed in the double ionization of Mg \cite{Gillen-etal} and several molecules even with circular polarization \cite{Guo-etal}. This is some support for the suggestion made above that the three curves in Fig. \ref{fD.3curves} show departures from ADK-type predictions.

These results are preliminary and not conclusive, but they represent an end-of-pulse examination of double ionization under ellipticities. They indicate that the full range of ellipticities from linear to circular polarization can be analysed using our purely classical ensemble method. The novel four-band $p_y$ structure identified for $\varepsilon$ = 0.5 also appears in preliminary results for other values of $\varepsilon$. We plan to extend these calculations, and expect on the basis of previous success with the classical model that prominent features appearing, such as the four-peak character of Fig. \ref{fD.4bands}, will be able to be explained by a straightforward trajectory analysis.

\end{document}